\documentclass{optica-article}

\journal{opticajournal}

\articletype{Research Article}

\usepackage{lineno}
\usepackage{tabularx}
\usepackage{xcolor}
\usepackage{soul}
\setstcolor{green}
\setulcolor{red}

\begin{document}


\title{Two-photon microscopy through scattering media harnessing speckle autocorrelation}

\author{Lei Zhu,\authormark{1} Bernhard Rauer,\authormark{1} Hilton B. de Aguiar, \authormark{1} and Sylvain Gigan\authormark{1,*}}

\address{\authormark{1}Laboratoire Kastler Brossel, ENS--Université PSL, CNRS, Sorbonne Université, College de France, 24 Rue Lhomond, F-75005 Paris, France.}

\email{\authormark{*}sylvain.gigan@lkb.ens.fr} 


\begin{abstract}
Two-photon (2P) microscopy is a powerful technique for deep-tissue fluorescence imaging; however, tissue scattering limits its effectiveness for depth imaging using conventional approaches. Despite typical strategies having been put forward to extend depth imaging capabilities based on wave-front shaping (WFS), computationally recovering images remains a significant challenge using 2P signal. In this work, we demonstrate the successful reconstruction of fluorescent objects behind scattering layers using 2P microscopy, utilizing the optical memory effect (ME) along with the speckle autocorrelation technique and a phase retrieval algorithm. Our results highlight the effectiveness of this method, offering significant potential for improving depth imaging capabilities in 2P microscopy through scattering media.
\end{abstract}

\section{Introduction}

Fluorescence-based microscopy techniques have evolved from a custom tool to a broadly available imaging modality in the life sciences. However, the transport of light through an inhomogeneous media, such as biological tissues, leads to the scattering phenomenon  \cite{bertolotti_imaging_2022,gigan_roadmap_2022}. In an imaging system, this scattering effect causes photons to deviate from the focal point when passing through a scattering medium. This effect significantly reduces the energy at the focal plane and introduces aberrations into the imaging system. Modern techniques, such as optical coherence tomography (OCT) \cite{abramson_light_flight_1978,huang_optical_1991}, actively reject scattered light while preserving ballistic light but are limited to thin or semi-transparent samples due to exponential signal loss with depth.
Another approach is the WFS technique, which compensates for scattering effects by focusing the scattered light to a diffraction-limited point \cite{vellekoop_focusing_2007,hsieh_imaging_2010,katz_looking_2012}. This is accomplished by optimizing the wavefront with a spatial light modulator (SLM), using an appropriate feedback signal from the system. However, this technique requires access to both sides of the scattering medium \cite{vellekoop_focusing_2007} or the presence of a guide star (or a known object) \cite{hsieh_imaging_2010,katz_looking_2012,horstmeyer_guidestar-assisted_2015}. Measuring the optical transmission matrix (TM) \cite{popoff_measuring_2010} enhances the ability to manipulate the output optical field by establishing a relationship between the input and output of the medium. Nevertheless, in practical biological applications, the use of the TM is limited by time-consuming acquisition and physical access constraints.

Recently, WFS combined with computational optimization methods was shown to enable fluorescence imaging through opaque samples. This has been achieved either by computationally retrieving the incoming and outgoing TMs \cite{boniface_non-invasive_2020,darco_physics-based_2022,weinberg_noninvasive_2024}, or by optimizing the different feedback signals to apply a physical correction to the excitation or detection path \cite{yeminy_guidestar-free_2021,aizik_fluorescent_2022}. Alternatively, retrieving either the incoming or outgoing TM allows focusing and imaging behind a complex medium\cite{zhu_large_2022,baek_phase_2023}. 

In non-linear optical fluorescence imaging scenarios, computational methods \cite{katz_noninvasive_2014,papadopoulos_scattering_2017,zhao_single-pixel_2024} have been proposed to address this challenge. These methods \cite{katz_noninvasive_2014,papadopoulos_scattering_2017,zhao_single-pixel_2024} effectively achieve focusing through scattering samples, followed by raster scanning to image within the range of optical ME \cite{bertolotti_imaging_2022,gigan_roadmap_2022}. However, refocusing through opaque samples using a spatial light modulator (SLM) remains relatively slow, constrained by both the computational speed of the algorithm and the modulation rate of the SLM itself.

Due to the presence of the optical ME, computational recovery of hidden objects behind complex samples has become widely used in fluorescence-based image reconstruction \cite{bertolotti_non-invasive_2012,katz_non-invasive_2014}. Here, we present a method for non-invasive imaging of hidden objects in a 2P fluorescence scenario using unknown random illuminations, without the need for a SLM. Leveraging the optical ME, the illumination laser pattern remains highly correlated during raster scanning, allowing the captured 2P fluorescence image to be expressed as the convolution of the scanning pattern and the hidden object. By applying an autocorrelation technique, the Fourier amplitude of the hidden object is preserved, enabling image reconstruction through an iterative phase retrieval algorithm. However, in 2P scenarios, the limited number of effective speckle grains and illumination power makes a single illumination realization insufficient for autocorrelation-based recovery. To overcome this limitation, multiple random speckle illumination realizations are used, ensuring a high signal-to-noise ratio (SNR) in the autocorrelation process. Finally, we experimentally demonstrate the applicability of our technique for imaging a hidden object through a random scattering medium in 2P microscopy.

\section{Principle}
In laser scanning microscopy, an image, $I$, is acquired by raster scanning a focused beam across the sample, $O$. The spatial profile of the laser beam, $S$, act as the point spread function (PSF) of the imaging system. The recorded image in 1-photon, $I_{1P}$, can be expressed as:
\begin{equation}
I_{1P}(\theta ) = \int{S(r) O(r-\theta d) d^{2}r} = [S\ast O](\theta )
 \label{EQ.1}
\end{equation}
where $\ast$ refers to the convolution operator, $\theta$ denotes the scanning angle, $d$ indicates the focus length of the imaging system, and $r$ represents the spatial coordinate.

\begin{figure}[ht!]
\centering\includegraphics[width=12cm]{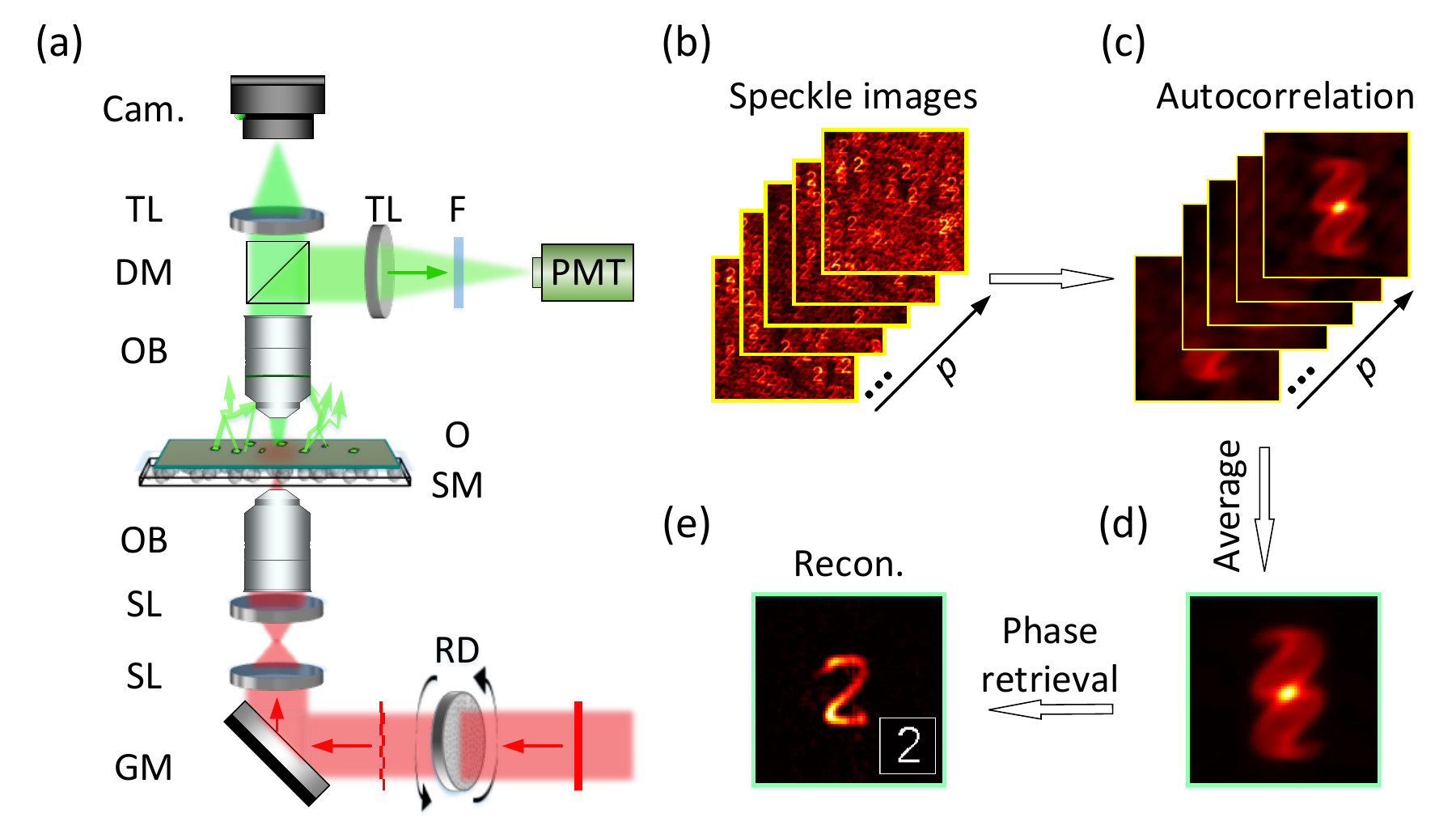}
\caption{\textbf{Imgaing through scattering media using 2P signal based on optical ME: concept and numerical example.} \textbf{a}, The laser beam passes through a rotating diffuser (RD), which randomly modulates the beam. The modulated beam is then delivered through the galvo mirror (GM) unit, which performs raster scanning, before reaching the back focal plane of the objective (OB). Then, the hidden object behind the scattering medium (SM) is excited. Upon excitation, the fluorescence passes through a second OB and is collected by a photomultiplier tube (PMT). \textbf{b}, A stack of speckle images is recorded by the PMT. \textbf{c}, The autocorrelation of each speckle. \textbf{d}, The averaged autocorrelation from (c). \textbf{e}, The final reconstruction is retrieved using an iterative phase-retrieval algorithm. SL: scanning lens, O: object, DM: dichroic mirror, TL: tube lens, F: filter, Cam.: camera, and Recon.: reconstruction.}
\label{fig.1}
\end{figure}

An ideal imaging system has a PSF that is a diffraction-limited Airy function, with its energy concentrated in the central lobe. On the contrary, a scattering media will scramble the phase, and distort the diffraction-limited spot, creating a random speckle pattern. Despite being spatially random, this speckle pattern still preserves high correlation while shifting within the optical ME range. Thus, the speckle pattern can still function as a PSF, and the convolution model remains valid in the speckle scenario within the optical ME range. However, 2P microscopy owes its advantages of nonlinear excitation, where the PSF changes to $|S|^2$, which leads to the suppression of weak speckle grains and to an enhanced contrast. In the 2P scattering imaging system, the captured 2P image, $I_{2P}$, can be denoted as 
\begin{equation}
I_{2P}(\theta ) = \int{|S(r)|^{2} O(r-\theta d) d^{2}r} = [|S|^{2} \ast O](\theta)
\label{EQ.2}
\end{equation}
Taking the autocorrelation of $I_{2P}$ and using the convolution theorem yields
\begin{equation}
I_{2P} \star I_{2P} = [|S|^{2} \ast O] \star [|S|^{2} \ast O]  = [O \star O] \ast [|S|^{2} \star |S|^{2}]   
\label{EQ.3}
\end{equation}
where $\star$ is the autocorrelation operator.

Within the ME range, the autocorrelation of $|S|^{2}$ is a sharply peaked function and its width is equal to the average speckle grain size. Then, the autocorrelation of the hidden object can be retrieved from the autocorrelation of the $I_{2P}$:
\begin{equation}
O\star O \approx (I_{2P}\star I_{2P}) +C
\label{EQ.4}
\end{equation}
$C$ is an additional constant background term. The autocorrelation of the hidden object, located behind various scattering medium, can be retrieved. 
Importantly, compared with 1P fluorescence experiments, due to the 2P effect, $I_{2P}$ doesn't contain enough efficient speckle grains to ensure a high SNR in the autocorrelation process when using a single frame. Therefore, multiple frames are acquired, and the average autocorrelation across these frames is used. Therewith, Eq.\ref{EQ.3} can be rewritten as:
\begin{equation}
O\star O \approx \langle I_{2P}\star I_{2P} \rangle  +C
\label{EQ.5}
\end{equation}
where $ \langle \cdot \rangle$ refers to the averaging operator.
According to the Wiener-Khinchin theorem, the Fourier transform of the autocorrelation gives the modulus of the object’s Fourier transform. 

\begin{equation}
|F\{O\}| \approx \sqrt{|F\langle I_{2P}\star I_{2P} \rangle|}
\label{EQ.6}
\end{equation}
Thus, the object’s image can be retrieved from its autocorrelation using a phase retrieval algorithm.

\section{Experimental setup and results}
The experimental scheme is shown in Fig. \ref{fig.1}a. A typical 2P microscopy setup, combined with an additional rotating diffuser (RD), is employed in our experiment. The laser beam, originating from an optical parametric oscillator (OPO, Coherent Mira OPO-X) pumped by a pulsed Ti-sapphire laser (Coherent Chameleon Ultra II) with a $140\ \mathit{fs}$ pulse duration and an $80 \ \mathit{MHz}$ repetition rate, passes through the RD, generating a randomly modulated light field. This modulated light then propagates through an optical scanning unit, consisting of two galvanometric mirrors (GM) unit and a 4f optical configuration with two scanning lenses (SL), ultimately reaching the back focal plane of a water-immersion objective (Zeiss W Plan-Apochromat 40×/1.0 DIC M27). The 2P excitation wavelength is centered at $1050 \ nm$.
Upon reaching the sample plane, the excitation light interacts with a scattering medium, positioned between the objective and the sample. The rough surface of the scattering medium is created by sandblasting the bottom surface of a coverslip. Fluorescent beads (Fluospheres carboxylate, $ 1.0\ \mu m$, $580\ nm$/\ $605\ nm$) are used as the samples. To acquire a single image, the RD is held in a fixed position while the excitation speckle is scanned over the sample by the GM.
A second objective (Zeiss EC Plan-NEOFLUAR 40×/1.3), placed opposite the first objective, collects the emitted fluorescence signal. The collected light is then diverted by a dichroic mirror (Semrock Di03-R785-t1-25 × 36), filtered by an optical filter to isolate the 2P fluorescence signal, and subsequently detected by a photomultiplier tube (PMT, Hamamatsu H7422P-40). Additionally, the second objective is used to image the object onto a camera (Basler acA1300-30 $\mu m$).

\begin{figure}[ht]
\centering\includegraphics[width=12cm]{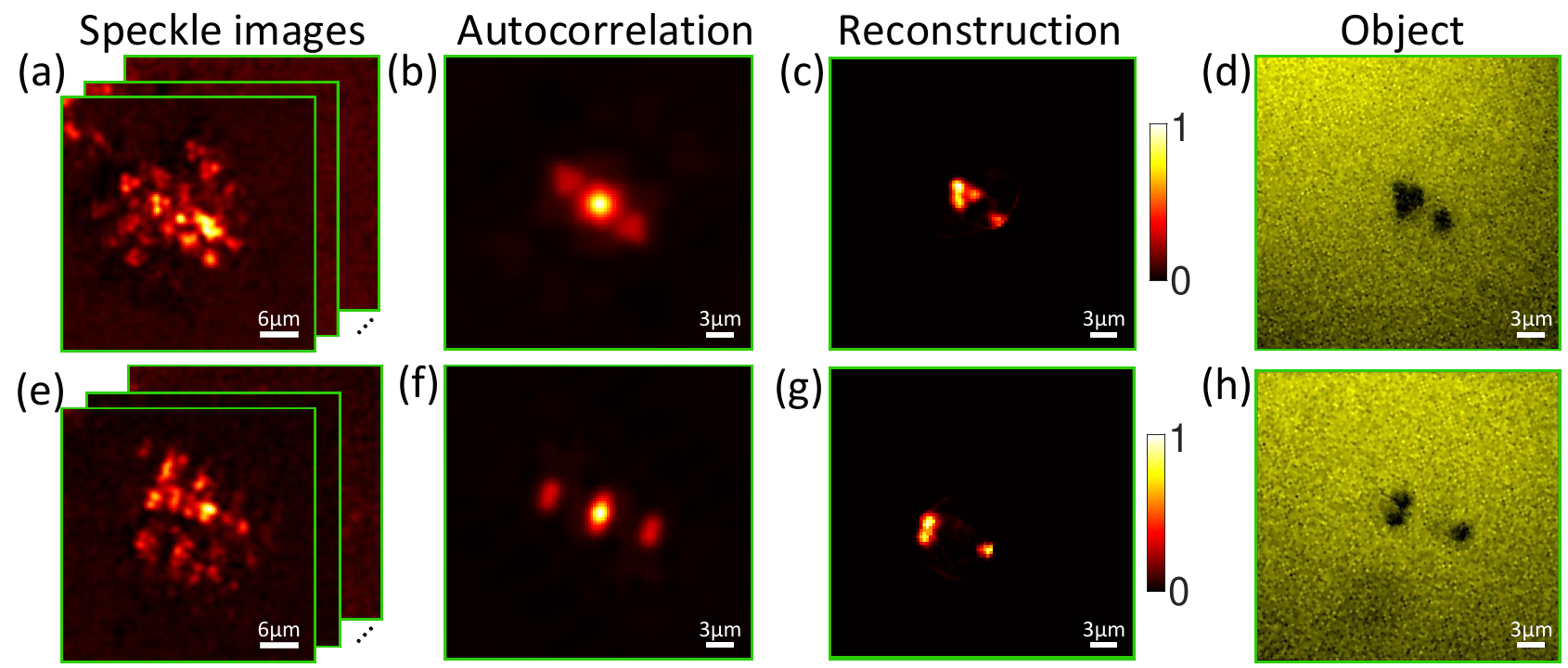}
\caption{\textbf{Experimental imaging through scattering media using 2P fluoresence signal based on optical ME.} \textbf{a,} The measured speckle images of 2P fluoresence signal. \textbf{b,} The average autocorrelation of the speckle images (a). \textbf{c,} The reconstructed object from the average autocorrelation in (b) by a phase retrieval algorithm. \textbf{d,} The image of the hidden object is taken without scattering medium. \textbf{e-h}, As in \textbf{a-d} but for different hidden object.}
\label{fig.2}
\end{figure}

The recorded $I_{2P}$ are presented in Fig. \ref{fig.2}a, where multiple images correspond to multiple spatial modulations of the illumination laser beam, obtained by rotating the diffuser. 
These different realizations are then used to average their autocorrelation $\langle I_{2P}\star I_{2P} \rangle$, as illustrated in Fig. \ref{fig.2}b. The different speckle realizations are independent using random unknown modulation from the RD. Therefore, comparing the measured images with respect to each other, such as cross-correlation, does not yield any information akin to the autocorrelation of individual images. Furthermore, the autocorrelation of a speckle is always centred and centrosymmetric. Notably, the average autocorrelation is offset-corrected by subtracting its minimum value.     

In order to retrieve the Fourier phase of the hidden object, we employ a combination of Fienup’s error reduction algorithm (ER) and a modified hybrid-input output algorithm (HIO) \cite{J1982Phase,bertolotti_non-invasive_2012,katz_non-invasive_2014,hofer_wide_2018,tian_single-shot_2022}. After retrieving the Fourier phase of $O$, the reconstruction is achieved by simply applying an inverse Fourier transform to the combined Fourier phase and amplitude of $O$, as presented in Fig. \ref{fig.2}c. In our work, a standard phase retrieval algorithm is employed, completed within a few seconds on a standard computer. In Fig. \ref{fig.2}d we display the image of the hidden object, taken without scattering medium in transmission geometry under white-light illumination. The reconstructed image in Fig. \ref{fig.2}c and the ground truth in Fig. \ref{fig.2}d show an excellent resemblance to each other. In Fig. \ref{fig.2}e-h, we present another sample realization to prove the effectiveness of our method in 2P fluorescence imaging. For the results presented in Fig. \ref{fig.2}a and \ref{fig.2}e, each experiment involves 80 speckle images, with each speckle image size being 128 × 128 pixels.

During the image reconstruction process, several parameters have to be carefully chosen. First, the recorded raw images were normalized using a Gaussian low-pass filtered version of themselves to remove the intensity envelope caused by the scattering process. The cutoff frequency of the low-pass filter was adjusted to match the object's dimensions, which were estimated based on the size of the autocorrelation. Second, a cosine window was applied to truncate the outer regions of the autocorrelation, which predominantly contain residual noise resulting from weak correlations of the object's replicas, beyond the optical ME range. These steps improve the convergence of the phase-retrieval algorithm. Third, the autocorrelation image was offset-corrected by subtracting its minimum value. An alternative approach can also be employed to separate the autocorrelation signal from background noise. By leveraging the sparseness of the autocorrelation and the low-rank nature of the background, Robust Principal Component Analysis (RPCA) \cite{bouwmans_robust_2014} enables the separation of the autocorrelation signal from its background noise \cite{li_lensless_2022}.

In the linear fluorescence scenario, the longitudinal position of the hidden object is not crucial and can be placed outside the focal plane of the illumination objective. However, in the case of 2P fluorescence, precise positioning of the hidden object becomes essential because the sectioning effect of 2P fluorescence causes the signal intensity to decrease sharply as the object moves out of focus, eventually disappearing entirely. A finite optical ME range, which combines optical shift and tilt ME \cite{osnabrugge_generalized_2017}, is often more practical for biological imaging applications. In our experiment, by positioning the scattering medium near the objective's focal plane and placing the RD in the equivalent back focal plane, we achieve a finite optical ME range.
 
\section{Conclusion and discussion}

We have demonstrated both numerically and experimentally the effectiveness of the proposed method for imaging a hidden object behind a scattering medium using 2P fluorescence signals. First, the object's size within the focal plane of the 2P imaging system must remain confined to the optical ME range. The extent of this ME range \cite{osnabrugge_generalized_2017} in the 2P imaging configuration is influenced by factors such as the system's numerical aperture (NA), the object's depth within the medium, and the optical properties of the scattering medium. While the optical ME range in biological tissues is inherently limited, our method shows significant potential for multiphoton biological imaging. Second, the 2P signal was collected using a forward-detection scheme with a high-NA objective to maximize signal collection and validate the principle. Notably, this experiment could also be implemented using an epi-detection scheme. Third, reconstructing the hidden object from captured speckle images simplifies the imaging process by eliminating the need to revisit the sample for actions like refocusing light. Moreover, the inherent optical sectioning capability of multiphoton imaging enables the reconstruction of 3D objects \cite{liu_3d_2015,mukherjee_3d_2018,soldevila_functional_2023} behind scattering samples through Z-range scanning. This reconstruction method can also be extended to three-photon fluorescence scattering imaging.

Despite these advantages, reconstructing more complex hidden objects remains challenging due to the limitations of phase retrieval algorithms. Other computational techniques also hold potential for enhancing multiphoton imaging applications. In particular, the excitation pattern in multiphoton fluorescence imaging exhibits higher-order properties compared to that of linear fluorescence imaging. Developing tailored strategies to leverage these unique excitation properties for hidden object reconstruction represents a promising avenue for future research.

\section{Funding}
This work was supported by FET OPEN DYNAMIC Chan Zuckerberg Initiative Deep Tissue Imaging; Agence Nationale de la Recherche (ANR-21-CE42-0013); H2020 Future and Emerging Technologies (863203); European Union’s Marie Skłodowska-Curie fellowship, grant agreement No. 888707 (DEEP3P).  S.G. is supported by Institut Universitaire de France.

\section{Acknowledgment}
The authors thank Fernando Soldevila and Payvand Arjmand for the careful reading of the manuscript.

\section{Disclosures}
The authors declare no conflicts of interest.

\section{Data availability}
Data underlying the results presented in this paper are available from the corresponding author upon reasonable request.

\bibliography{main}

\end{document}